\documentclass[12pt]{article}
\usepackage{authblk}
\usepackage{amsmath,amssymb}
\usepackage{listings}

\usepackage{amsthm}
\usepackage{tikz-qtree}
\usepackage{tikz-qtree-compat}

\newtheorem*{lemma}{Lemma}

\tikzset{every tree node/.style={align=center, anchor=north}}

\makeatletter
\renewcommand\@biblabel[1]{#1.}
\makeatother

\begin{document}

\title{A fast algorithm for constructing balanced binary search trees}
\author[]{Pavel S. Ruzankin\footnote{ruzankin@math.nsc.ru}}
\affil[]{\normalsize Sobolev Institute of Mathematics, Novosibirsk, Russia\\
Novosibirsk State University, Novosibirsk, Russia}
\date{}
\maketitle
\begin{abstract}
We suggest a new non-recursive  algorithm for constructing a binary search tree given an array of numbers. The algorithm has $O(N)$ time and $O(1)$ memory complexity if the given array of $N$ numbers is sorted. The resulting tree is of minimal height and can be transformed to a complete binary search tree (retaining minimal height) with $O(\log N)$ time and $O(1)$ memory. The algorithm allows simple and effective parallelization. 

\end{abstract}

A binary search tree (BST) is a fundamental data structure which is widely used in applications.
There is a large variety of algorithms for constructing BST's. 
The first approach is based on sequentially adding nodes to the tree. The nodes may be 
added to leaves \cite{knuth} or to the root of the tree \cite{stephenson}.
The second approach consists in reconstructing a BST from preorder or
postorder traversals (e.g., see \cite{kopp,das} and references therein). 
The third approach is based on halving the given sorted array and recursively building  
the left and the right subtrees \cite{wirth}. There are also algorithms that account for
the probabilities of hitting specific nodes and try to build optimal BST's 
(e.g., see \cite{gagie} and references therein).

There also exist algorithms that do not adhere to those approaches \cite{rsci, vaucher}.
The recursive algorithm in \cite{vaucher} constructs the tree by sequentially constructing perfect 
BST's. After a perfect BST is constructed, it is 
incorporated into the new BST as the left subtree of the root. Then the new BST is 
built up to a perfect BST, and so on.

We present a new non-recursive algorithm
for constructing a binary search tree.
The algorithm has $O(N)$ time and $O(1)$ memory complexity if the given array of $N$ numbers is
sorted. We use an array-based representation of  the BST. The $O(1)$ memory complexity means that, except for 
the resulting arrays used to store the tree, we need $O(1)$ memory.
If the link-based representation is needed
then the algorithm will additionally need $O(N)$ memory.
The resulting BST has the minimal height, though may not be balanced in the sense of AVL trees, i.e.,
the trees where the heights of the two child subtrees of each node differ by at most one.
The new algorithm, though being non-recursive, 
somehow resembles the recursive algorithm in 
\cite{vaucher}. Moreover, we can use the rotations algorithm from \cite{vaucher}
to make the BST complete, retaining the minimal height, which needs $O(\log N)$ time
and $O(1)$ memory.

We will assume that we have already sorted the given array of $N$ numbers.
To simplify notations, we will build a BST for the numbers $0,...,N-1$.

Our algorithm is substantially based on the binary representation of a number. We will mark the binary numbers with a leading zero to distinguish them from decimal ones; e.g., $2=010$.

\paragraph*{}
First, let us consider the case when $N=2^K-1$ for some integer $K\ge 1$.
In this case the minimal height BST is perfect. For example, for $K=4$, the tree is shown on Fig.~1. 

\begin{figure}
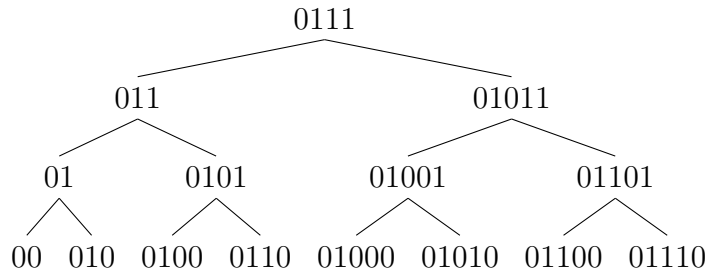

\begin{center}
\Tree[.0111 [.011 [.01 00  010 ] [.0101 0100  0110 ]]
          [.01011 [.01001 01000 01010 ] [.01101 01100 01110 ]]]
\end{center}
\caption{An example of a perfect binary search tree}
\end{figure}


Let the {\it level} of a node  be the distance from the node to the nearest leaf
in the perfect BST. That is, in a perfect BST, the leaves lie on level 0, the parents of the leaves lie on level 1, and so on. Note that the level of a node depends on the number of the node only and does not depend on the height of the tree.

We see that the binary representations of nodes of level $k$ end with $k$ ones leaded by zero, since subsequent nodes on level $k$ differ by $2^{k+1}$.
Thus the level $L(j)$ of a node $j$ can be calculated as the location of the least significant zero in the binary representation of $j$.

To calculate $L(j)$, one may use the operation of the least significant one in a binary number which is implemented in many modern processor architectures \cite{ffs}. There are also 
built-in functions for the operation in popular compilers.
For instance, in GCC, $L(j)$ can be defined as 
 \verb|__builtin_ffs(~j)-1|. We assume that the binary representation of $N-1$ 
 contains at least one zero, which is the case when 
 $N$ can be represented as the number of the same unsigned integer type as is used for indexing the cells of the given sorted array
 of numbers.

However Algorithms 1 and 2 below utilize not $L(j)$ itself but $2^{L(j)}$, and Algorithm 3 below can be obviously modified to 
use $2^{L(j)}$ instead of $L(j)$ if needed. 
It is well known \cite{ffs} that 
$$2^{L(j)}=(j+1) \& (-(j+1))$$
or 
$$2^{L(j)}=(\sim j) \& (-(\sim j)),$$
where $\&$ is the bitwise AND operator, $\sim$ is the bitwise NOT operator, $-j$ is the negative of j treating j as a signed integer in two's complement arithmetic which is common in modern processors. For instance, in R, $2^{L(j)}$ can be defined as
\verb|bitwAnd((j+1),-(j+1))|.

A node $j$ on level $k\ge 1$ has the left child $j-2^{k-1}$ and the right child
$j+2^{k-1}$ . Besides, a node $j$ on level $k\ge 0$ has the parent $j+2^k$ if the
binary representation of $j$ ends with ``$001...1$'' (ending with $k$ ones)  and the parent $j-2^k$
if the binary representation of $j$ ends with ``$101...1$''.

Thus, for the case $N=2^K-1$,  we can write down the algorithm as the following pseudocode.
Below $p$, $l$, $r$ denote the resulting arrays of parents, left children, and right children, respectively.
The algorithm constructs the BST as these three arrays. $M(j)$ denotes the location of the most significant one in the binary representation of $j$, e.g., $M(01001)=3$;
$t$ is the number of the root node. 

\newpage
\noindent{\bf Algorithm 1.}
\begin{lstlisting}
for ($j$ in $0,\dots, N-1$)
   if ($(j\ \&\ 2^{L(j)+1})=0$)
      $p$[$j$]:=$j+2^{L(j)}$
   else
      $p$[$j$]:=$j-2^{L(j)}$
   end if
   if ($L(j)>0$)
      $l$[$j$]:=$j-2^{L(j)-1}$
      $r$[$j$]:=$j+2^{L(j)-1}$
   else
      $l$[$j$]:=NULL
      $r$[$j$]:=NULL
   end if
end for
$t$:=$2^{M(N)}-1$
$p$[$t$]:=NULL
\end{lstlisting}

\paragraph*{}
Now it remains to modify Algorithm~1 for the case of arbitrary $N$. 
If we try to build a binary tree with Algorithm~1 then some edges may point
to missing nodes that are greater than $N-1$.
\begin{lemma}
All the edges pointing to missing nodes in the ``tree'' built by Algorithm~1, except 
 the down-right edge of the last node if any, are located on the ascending path from
 the node $(N-1)$
to the root in the perfect BST of the same height. 
\end{lemma}
\proof
Let us have the ``tree'' constructed by Algorithm~1. 
Let a node $j$, $j\ne N-1$, of the ``tree''
have its down-right edge pointing to a missing node $i$.
We have $j<N-1<i$. Let $k$ be the level of the node $j$,
and let $m$ be the ancestor of the node $(N-1)$ on level $k$
in the corresponding perfect BST.
Then $m\ge j$ since $j<N-1$. Besides, we cannot have $m>j$ since
it would imply $N-1>i$. Hence, $m=j$, and the ancestor
of the node $(N-1)$ at level $k-1$ in the corresponding perfect BST is $i$
since $j<N-1<i$.

Let now a node $j$ of the ``tree'' constructed by Algorithm~1
have its up edge pointing to a missing node $i$,
let $k$ be the level of the node $j$,
and let $m$ be the ancestor of the node $(N-1)$ on level $k$
in the corresponding perfect BST.
Then again $j<N-1<i$ and $m\ge j$, and again 
we cannot have $m>j$ since
it would imply $N-1>i$. Hence, $m=j$.

The lemma is proved.

\medskip To correct the ``tree'' built by Algorithm~1,
it remains to follow the descending path from
the root to the node $(N-1)$
in the corresponding perfect BST 
and ``glue'' edges pointing to missing nodes.
Finally, the algorithm is as follows.  Below $/$ denotes integer division, e.g., $1/2 = 0$.

\medskip
\noindent{\bf Algorithm 2.}
\begin{lstlisting}
function $P(j)$ := 
   if ($(j\ \&\ 2^{L(j)+1})=0$) then $j+2^{L(j)}$
   else $j-2^{L(j)}$

for ($j$ = $0$ to $N-1$ by $2$)
   $p$[$j$]:=$P(j)$
   $l$[$j$]:=NULL
   $r$[$j$]:=NULL
end for
for ($j$ = $1$ to $N-1$ by $2$)
   $p$[$j$]:=$P(j)$
   $l$[$j$]:=$j-2^{L(j)-1}$
   $r$[$j$]:=$j+2^{L(j)-1}$
end for
$r$[$N-1$]:=NULL
$t$:=$2^{M(N)}-1$
$p$[$t$]:=NULL

$k$:=$2^{L(t)}$
$j$:=$t$
while ($k> 2^{L(N-1)} $)
   $k$:=$k/2$
   if ($((N-1-t)\ \&\ k)=0$)
      $k$:=$k/2$
      while ($((N-1-t)\ \&\ k)=0$)
         $k$:=$k/2$
      end while
      $r$[$j$]:=$j+k$
      $p$[$j+k$]:=$j$
   end if
   $j:=j+k$
end while
\end{lstlisting}

\medskip
The time complexity is still $O(N)$, since ``gluing'' edges after
the \verb|for| loops takes $O(\log N)$ time and $O(1)$ memory.

The \verb|while| loops for ``gluing'' edges are explained as follows. Traveling by the descending path from the root node $t$ to the node $(N-1)$, we move by $\pm 2^{L(j)-1}$ when we go from the node $j$ to its right/left child. Thus, $N-1-t=A-B$, where $A$ is the binary number with 1's on locations $m$ such that the path contains the edge from $j$ to its right child, where 
$m=L(j)-1$. Analogously, $B$ has 1's on locations $m$ such that the path contains the edge from $j$ to its left child, 
$m=L(j)-1$. The path goes through the nodes $>N-1$ when it contains a subpath with the edges ``right--left--left--\dots--left'' with
the next edge being down-right or with the last edge of the subpath being the last edge of the path. 
Only the first and the last node of the subpath are $\le N-1$.
So we must ``glue'' each such subpath into one edge. Let $m$ and $n$ be the levels of the first and the last node of the subpath. Then $N-1-t$ will contain the following binary digits at the locations $m-1,\dots,n$: $100\cdots0-011\cdots1=00\cdots01$. The \verb|while| loops just search for all such subpaths (all such patterns in $N-1-t$) and connect their first and last nodes with an edge.

\medskip
{\bf Remark 1.} Algorithm 2 allows simple and effective parallelization. The only loop that cannot be parallelized is the loop 
correcting edges pointing to nodes $>N-1$. That loop has complexity $O(\log N)$.

\medskip
{\bf Remark 2.} 
Algorithm 2 can be used without really constructing the tree. In this case the
tree is ``virtual'', we need no time and no memory to construct the tree; the search operation needs $O(\log N)$
time, and the number of examined nodes for each search does not exceed the (minimal) height; the operations of
deletion and insertion of nodes are just the deletion and insertion of a number to the given array keeping it sorted.
The search operation for the ``virtual'' tree is defined as follows. A search path is the path in the corresponding perfect
BST such that when we meet a node $>N-1$ we go down-left until we reach some node $\le N-1$. 

\medskip
{\bf Remark 3.} If a user does not need the array of parents $p$ then the array
can be excluded from Algorithm~2 as well as from Algorithm~3 below, since those algorithms do not read the values from $p$.

\paragraph*{}
To make the tree complete (retaining the minimal height) we can use the rotations algorithm from 
\cite{vaucher} as follows.

\medskip
\newpage
\noindent{\bf Algorithm 3.}
\begin{lstlisting}
function $R(j)$:= 
   if ($j=N-1$) then $0$
   else $M(N-1-j)+1$

$x$:=$t$
$h$:=$L(t)$
if ($R(x)<h$ and $h> 1$) then
   $y$:=$l$[$x$]
   $t$:=$y$
   $p$[$y$]:=NULL
   $l$[$x$]:=$r$[$y$]
   $p$[$l$[$x$]]:=$x$
   $r$[$y$]:=$x$
   $p$[$x$]:=$y$
   $z$:=$y$
else 
   $z$:=$x$
   $x$:=$r$[$x$]
end if 
$h$:=$h-1$

while ($h> 1$)
   if ($R(x)<h$) then
      $y$:=$l$[$x$]
      $p$[$y$]:=$z$
      $r$[$z$]:=$y$
      $l$[$x$]:=$r$[$y$]
      $p$[$l$[$x$]]:=$x$
      $r$[$y$]:=$x$
      $p$[$x$]:=$y$
      $z$:=$y$
   else 
      $z$:=$x$
      $x$:=$r$[$x$]
   end if 
   $h$:=$h-1$
end while
\end{lstlisting}

Here $R(j)$ stands for the height of the right subtree of a node $j$ in the BST built by Algorithm~2 when
the node $j$ is reachable from the root node by descending via down-right edges only,
$h$ is the level of the current node, $x$ is the current node. Algorithm~3 needs $O(\log N)$ time since
it goes down by $1$ in $h$ each iteration of the \verb|while| loop.


\begin{thebibliography}{99}
%

\bibitem{kopp}
N. Aghaieabiane, H. Koppelaar, P. Nasehpour (2017). An improved algorithm to reconstruct a binary tree from its inorder and postorder traversals.  Journal of algorithms and computation.  vol. 49, no. 1, pp. 93--113.

\bibitem{das}
V. V. Das (2010). A New Non-recursive Algorithm for Reconstructing a Binary Tree from its Traversals. 2010 International Conference on Advances in Recent Technologies in Communication and Computing, Kottayam, pp. 261--263.
doi: 10.1109/ARTCom.2010.88

\bibitem{ffs}
Find first set. Wikipedia article. retrieved January 14, 2019. https://en.wikipedia.org/wiki/Find\_first\_set

\bibitem{gagie}
T. Gagie (2003). New Ways to Construct Binary Search Trees. In: Ibaraki T., Katoh N., Ono H. (eds) Algorithms and Computation. ISAAC 2003. Lecture Notes in Computer Science, vol. 2906, Springer, Berlin, Heidelberg.

\bibitem{knuth} 
D. E. Knuth (1973). The art of computer programming: Sorting and searching., vol. 3.
Reading, Mass.: Addison-Wesley Pub. Co.

\bibitem{rsci}
Sorted List to complete BST.  Rhyscitlema.  July 31, 2018. retrieved February 22, 2019. http://rhyscitlema.com/algorithms/sorted-list-to-complete-bst/

\bibitem{stephenson}
C. J. Stephenson (1980).
A method for constructing binary search trees by making insertions at the root. Int. J. Comput. Inf. Sci. 9, pp. 15--29.

\bibitem{vaucher}
J. G. Vaucher (2004). Building optimal binary search trees from sorted values in O(N) time. In: Essays in Memory of Ole-Johan Dahl, pp. 376--388.

\bibitem{wirth}
N.  Wirth (1976).  Algorithms  +  data  structures = programs.  Englewood  Cliffs,  N.J.:
Prentice-Hall.
\end{thebibliography}
\end{document}